\documentclass[aps,twocolumn,superscriptaddress,showpacs,pre,floatfix]{revtex4-1}
\usepackage{graphicx,color}
\usepackage{amssymb}
\newcommand{\corr}[1]{{\leavevmode\color{black}#1}}
\usepackage{comment,color}

\begin{document}
\title{A multi-component model for the IceCube neutrino events}
\author{A.~Palladino} 
\affiliation{Gran Sasso Science Institute}
\email{andrea.palladino@gssi.infn.it}
\begin{abstract}
The IceCube neutrino telescope has observed for the first time a diffuse flux of high energy neutrinos, with a possible astrophysical origin. Up to now there are no evidence of sources and many hypothesis are still plausible in order to explain the measured flux. In this proceeding we analyze an alternative way to interpret the IceCube neutrinos, in terms of  sum of contributions from different sources.
\end{abstract}

\maketitle


\section{Introduction}
The IceCube neutrino telescope is a km$^3$ detector, located in the South Pole. It has provided the first evidence of a diffuse flux of high energy neutrinos (HE$\nu$), with a possible astrophysical origin \cite{ice2}. At the present the excess measured by IceCube, above the expected atmospheric background, has a significance greater than 6$\sigma$ \cite{ice3,ice4}.

The IceCube neutrino telescope observes neutrinos in two different ways: 

\textit{1) high energy starting events (HESE)}, \corr{mainly hadronic and/or electromagnetic showers from Southern hemisphere, with the vertex of interaction contained into the detector, above 30 TeV. HESE are sensible to all neutrino flavors.} 

\textit{2) throughgoing muons}. They are induced muons, produced by the charge current interaction of $\nu_\mu$ outside the detector. 
They come from the Northern hemisphere \cite{icemuon}.

Historically the diffuse flux of HE$\nu$ from extragalactic sources was expected to be distributed as $E^{-2}$ \cite{waxman}. 
At the present  the HESE suggest a softer spectrum $E^{-\alpha}$, with $\alpha=2.5 \pm 0.1$ \cite{ice4}. On the contrary, the dataset of 29 throughgoing muons \cite{icemuon}, with deposited energy above 200 TeV, is in agreement with a harder spectrum with $\alpha= 2.13 \pm 0.13$. There is a tension of 3.6$\sigma$ among them. \corr{On the other side the highest energy part of HESE suggests an $E^{-2}$ spectrum, in agreement with throughgoing muons.}

Here we discuss the possibility that the two hemisphere are observing different populations of HE$\nu$ \cite{gal1,gal2}.

\section{The different contributions}
\textbf{Extragalactic}: the throughgoing muons dataset is well representative of the extragalactic component of HE$\nu$, thanks to the high energy threshold. At this energy the contamination given by the conventional atmospheric background is negligible whereas there can be a certain contamination due to prompt neutrinos, produced into the decay of heavy mesons. In \cite{bllac} it has been estimated that 2/3 of the troughgoing muons events can be attributed to the astrophysical signal and 1/3 to the atmospheric background. 
Therefore it is reasonable to assume that the extragalactic flux of HE$\nu$ is distributed as $E^{-2}$ \cite{gal1} or as $E^{-2.13}$ \cite{gal2} \corr{and not as $E^{-2.5}$, as suggested by HESE \cite{ice4}.}

On the theoretical point of view, there are several sources that have the potential to produce HE$\nu$. Nowadays, despite the low statistic of IceCube, some theoretical scenarios are already in strong tension with the observations:

\textit{Gamma Ray Burst}. They are disfavored by the non observation of correlation in space and time with the IceCube neutrinos. Their contribution to the diffuse flux can be no more than few \% \cite{icegrb};

\textit{Blazars.} They are a subclass of AGNs with the emitting jet pointing into direction of the Earth. Blazars are the brightest objects in the $\gamma$-rays sky above 100 GeV. Therefore it is natural to consider them as a promising sources of high energy neutrinos. Nowadays it is known that \textit{1)} there are correlations between HESE and blazars \cite{resconiblaz}. Anyway HESE are mainly shower-like events, with a poor angular resolution of 10$^\circ$-15$^\circ$;
\textit{2)} the IceCube collaboration has calculated that the contribution of blazars to the diffuse flux of HE$\nu$ is not greater of 25\% \cite{iceblaz};
\textit{3)} blazars are strongly constrained by the non observation of multiplets \cite{murasebl,tamborrabl};
\corr{\textit{4)} a subclass of blazars, the BL Lacs, are strongly constrained by the non observation of correlations with throughgoing muons above 200 TeV \cite{bllac}.}

\textit{Starbust galaxies.} They are Galaxies in which the gas density is much higher than what is observed in quiescent galaxies and for this reason the proton-proton ($pp$) interaction is a plausible mechanism to produce HE$\nu$. At the present there is no tension between the flux of HE$\nu$ \corr{theoretically} expected 
and the flux detected by IceCube \cite{tamborramurase}.

\textbf{Galactic from disk}:
at the present a Galactic component of HE$\nu$ has been not measured. There is only an upper limit on this flux, provided by the ANTARES collaboration \cite{upperantares}. Anyway a diffuse flux of Galactic $\gamma$-rays from disk has been observed by Fermi \cite{fermigal} and it is plausible that also a Galactic flux of HE$\nu$ could exist.

The flux of Galactic neutrinos could be produced by the $pp$ interaction of Galactic cosmic rays with the matter contained into the Galactic disk. In this type of interaction neutrinos take about 5\% of the primary proton's energy and the spectrum of primary protons is replicated by neutrinos. 
Since the Galactic cosmic rays are distributed as $E^{-2.7}$ before the knee (at 3 PeV for cosmic rays), we expect that Galactic neutrinos are distributed as $E^{-2.7}$ and they give a contribution between 30 TeV and $\sim$ 150 TeV \cite{gal1}. On the other side, looking at the diffuse flux of Galactic $\gamma$-rays, an $E^{-2.4}$ spectrum for Galactic neutrinos seems to be more plausible, because an $E^{-2.7}$ spectrum is hard to reconcile with the Galactic $\gamma$-rays, when it is extrapolated at TeV energy\cite{gal2}.

\textbf{Prompt neutrinos:}
the atmospheric prompt neutrinos, produced in the decay of heavy mesons, are expected \cite{ers} but still not detected. Up to now there is an upper limit provided by IceCube \cite{icemuon}. The flux of \corr{these} neutrinos is expected to be distributed as $E^{-2.7}$, i.e. the same spectral index of the observed cosmic rays in the TeV-PeV range.
We do not expect a contribution of prompt neutrinos greater than few \% to the HE$\nu$ flux. Moreover it is important to take into account also this small effect that can \corr{contribute to} soften the spectrum below 100 TeV.

\section{The multi-component flux}
On the light of what said before, the diffuse flux of HE$\nu$ observed by IceCube can be explained as the sum of different contributions, as follows:
$$
\frac{d\phi}{dE_\nu} = \sum_{i=1}^3 N_i \times \frac{10^{-18} }{\rm GeV \ cm^2 \ sec \ sr} \left(\frac{E_\nu}{\rm 100 \ TeV} \right)^{-\alpha_i}
$$
where: -$N_1=2.7 \pm 0.9$ and $\alpha_1=2.13 \pm 0.13$ are the coefficients of the extragalactic component, \corr{that is isotropically} distributed and it is the dominant one. It should be responsible of $25 \pm 3$ of the 54 HESE observed; -$N_2=1.5 \pm 0.8$ and $\alpha_2=2.4$ are the coefficients of the Galactic component, that gives a smaller contribution. This component is present only in the flux that comes from the Southern hemisphere, in first approximation, and it produces $6.0 \pm 3.5$ events \cite{gal1,gal2};
-$N_3=0.6 \pm 0.3$ and $\alpha_3=2.7$ are the coefficients given by prompt neutrinos. This normalization respects the IceCube upper limit, namely 0.5 ERS \cite{icemuon} and it can produce only a small of fraction of HESE, namely $3.5 \pm 1.2$.  The other events are due to the conventional atmospheric background and to atmospheric muons (see \cite{gal2,ice3}). This model is compatible with the most recent constraints on the Galactic component, provided by ANTARES \cite{Albert:2017oba} and by IceCube \cite{Aartsen:2017ujz}.

\section{Conclusion}
The discovery of a diffuse flux of high energy neutrinos has opened a new era for neutrino astronomy. Up to now there are several theoretical models that are able to explain the observed flux and more data are required to clarify the situation. Most of the flux observed by IceCube is likely to have an extragalactic origin, but the source is unclear; GRBs and BL Lacs seem to be disfavored, Starburst Galaxies are still into the game, but at the present it is not possible to say much more.  \corr{A part} of the IceCube signal \corr{could be} given by a Galactic component, produced by the $pp$ interaction of Galactic cosmic rays with the matter contained into the Galactic disk. \corr{This hypothesis} can reconcile \corr{the spectral} tension observed in the IceCube data; \corr{moreover} a null Galactic component is disfavored at 2 sigma by spectral and spatial informations. 
Also the small contribution of atmospheric prompt neutrinos should be taken into account below 100 TeV. Therefore a multi-component model is reasonable to explain the IceCube data and it could be a good improvement respect to the single power law model.

\end{document}